\documentclass[pra,twocolumn,showpacs,superscriptaddress, floatfix]{revtex4-2}
\usepackage{epsfig,graphicx,amssymb,color,mathrsfs,amsthm}
\usepackage{amsmath}
\usepackage[colorlinks, linkcolor=blue, urlcolor=blue, citecolor=blue, unicode=true]{hyperref}
\begin{document}

%\begin{spacing}{1}
\title{Quantum master equation approach for the multiphonon up-pumping model}

\author{Jiong Cheng}
\affiliation{Department of Physics, School of Physical Science and Technology, Ningbo University, Ningbo 315211, China}
\author{Yanqiang Yang}
\affiliation{National Key Laboratory of Shock Wave and Detonation Physics, Institute of Fluid Physics, China Academy of Engineering Physics, Mianyang 621900, China}
\author{Wenlin Li}
\affiliation{College of Sciences, Northeastern University, Shenyang 110819, China}
\author{Xun Li}\thanks{lx\_phy@163.com}
\affiliation{National Key Laboratory of Shock Wave and Detonation Physics, Institute of Fluid Physics, China Academy of Engineering Physics, Mianyang 621900, China}
\date{\today }

\begin{abstract}
	A fully quantum multiphonon up‑pumping model is proposed to characterize coherent energy transfer in energetic materials (EMs) subjected to external shock.
	After eliminating the degrees of freedom of the phonon bath within a mean-field approximation, we derive a quantum master equation governing the energy transfer among vibrational modes.
	Our analysis reveals that doorway modes of different frequencies undergo distinct levels of effective coherent driving and dissipation, induced by the shocked phonon environment.
	This not only clarifies the microscopic origin of coherent phonon generation, but also reveals the possibility of modulating such coherent driving and dissipation.
	Based on numerical simulations of a simplified model using the master equation, we demonstrate how doorway modes extract energy from the phonon environment and subsequently excite higher-frequency molecular vibrational modes.
	This work offers a renewed perspective for understanding the mechanisms of energy transfer in energetic materials.
\end{abstract}

\maketitle

\section{Introduction}
\label{sec:Introduction}

The multiphonon up-pumping theory~\cite{dlott1986optical,hill1988model,dlott1990shocked,kim1990theory} elucidates the energy transfer mechanism in energetic materials under shock.
It posits that impact energy is initially converted into low-frequency phonons (lattice vibrations), which subsequently pump energy via multiphonon processes into high-frequency molecular vibrational modes (doorway modes).
This excitation of critical vibrational energy levels ultimately leads to chemical bond cleavage.
The theory bridges the gap between externally generated phonons and the onset of chemical reactions in such materials.

The proposal of the multiphonon up-pumping theory bridges energy transfer between localized intramolecular vibrations and delocalized lattice modes.
Subsequent studies~\cite{mcnesby1997spectroscopic,hooper2010vibrational,michalchuk2019predicting,bidault2022can,liu2024probing} have effectively advanced the understanding of the microscopic mechanisms of hotspot formation in energetic materials.
Research has shown that combining the multiphonon up-pumping model with classical phenomenological hotspot models effectively explains the hotspot formation mechanism during crystal grain fragmentation~\cite{mcnesby1997spectroscopic}.
Other researchers have applied the multiphonon up-pumping model in conjunction with shock wave theory to investigate the detonation initiation process of explosives under shock loading.
Based on the relationship between the excitation efficiency of doorway modes and the Debye frequency of crystals, a semi-classical theory of multiphonon up-pumping under shock loading was proposed~\cite{hooper2010vibrational}.
In recent years, a series of first-principles simulation studies have been conducted, such as analyzing the density of states, the number of doorway modes, and the matching relationship between doorway modes and specific vibrational modes frequencies in energetic material crystals~\cite{michalchuk2019predicting,bidault2022can}.
Additionally, the influence of phonon-vibration coupling coefficients on impact sensitivity has been calculated~\cite{liu2024probing}, leading to the establishment of correlations between phonon-related statistical measures and the impact sensitivity of explosives.

However, current research has largely overlooked critical aspects of post-loading phonon dynamics, including transport, scattering, and absorption. Additionally, the bidirectional nature of energy exchange between nonlinearly coupled vibrational modes remains unaddressed.
Furthermore, the potential influence of quantum coherence in vibration-vibration coupling on energy transfer efficiency has not been adequately considered, i.e. the quantum dynamics of this nonlinearly coupled multimodes regime named as multiphonon up-puming model of EMs has not been studied.
As a result, a microscopic description of hotspot formation of energetic materials under dynamic loading remains unattainable.

Addressing these limitations requires the development of a quantum theoretical framework, from a microscopic perspective, to explain the energy transfer mechanisms within the microstructure of energetic materials.
Given the continuous nature of the phonon spectrum, we can treat the phonon modes as the environment and the molecular vibrational modes as the system.
This approach allows for the proposal of a microscopic phonon bath and vibration-vibration coupling model based on the theory of open quantum systems~\cite{breuer2002theory}.

Research on the dynamics of open quantum systems gained significant attention during the 1970s and 1980s.
Most of these early works approached the subject from the perspective of mathematical physics~\cite{jamiolkowski1974effective,gorini1978properties}, with discussions largely focusing on the statistical properties of open quantum systems~\cite{alicki1979entropy}.
In recent years, this theoretical framework has found applications across interdisciplinary fields such as quantum chemistry, quantum biology, and quantum information science~\cite{rotter2015review}.
For instance, the quantum mechanisms underlying energy transfer in photosynthesis have been discovered~\cite{engel2007evidence}, and subsequently explained through studies based on open quantum system theory~\cite{ishizaki2012quantum}.
In the multiphonon up-pumping process, nonlinear couplings dominate, not only between the system and the environment but also among vibrational modes within the system itself.
Therefore, constructing physically sound models tailored to these nonlinear coupling characteristics and the specific features of energetic materials, along with developing corresponding numerical solution methods, is crucial for addressing this class of nonlinear open quantum system problems.

In this paper, we construct a fully quantum model of multiphonon up-pumping.
By treating the continuum spectrum of phonons below the Debye frequency as an environment and eliminating its degrees of freedom via a mean-field approximation, we derive a quantum master equation that describes the energy transfer among vibrational modes.
We find that the doorway modes experience both an effective coherent drive and effective dissipation induced by excitation from the phonon environment.
Since coherent phonons and thermal phonons with lost phase correlation interact with vibrational modes in markedly different ways, our findings provide microscopic evidence for the existence of coherent phonons.
Moreover, we identify key factors that determine the generation of coherent phonon driving, such as the initial number of phonon excitations from external shock, the phonon density of states, and the phonon-vibration coupling strength.
These factors determine that doorway modes at different frequencies experience varying strengths of coherent driving and dissipation.
Consequently, an efficient energy transfer process requires that the distribution of doorway modes within the energetic material satisfy specific matching conditions.
Subsequently, taking a simplified multiphonon up-pumping model as an example, we numerically simulate the energy transfer process involving five vibrational modes.
Our results illustrate how doorway modes draw energy from the phonon environment and further excite higher-frequency molecular vibrational modes.
This study offers a microscopic, fully quantum perspective on the multiphonon up-pumping theory, advancing the understanding of energy transfer mechanisms in energetic materials under external shock from the standpoint of quantum dynamics.

The paper is organized as follows.
In Sec.~\ref{sec:HSD}, we describe the Hamiltonian and derive the master equation.
In Sec.~\ref{sec:NR}, we simulate the master equation based on a simplified multiphonon up-pumping model.
Finally, we conclude in Sec.~\ref{sec:DC}.

%%%%%%%%%%%%%%%%%%%%%%%%%%%%%%%%%%%%%%%%%%%%%%%%%%%%%%%%%%%%%%%%%%%%%%%%%%%%%%%%%%%%
\section{Hamiltonian and system dynamics}
\label{sec:HSD}

The multiphonon up-pumping theory describes the mechanical (vibrational) excitation of the abundant nonlinear molecules in energetic materials using a Hamiltonian characterized by
\begin{eqnarray}
	\hat{H} &=& \hat{H}_{\text{ph}} + \hat{H}_{\text{vib}} + \hat{H}_{\text{ph-vib}} + \hat{H}_{\text{vib-vib}}, \notag\\
	\hat{H}_{\text{ph}} &=& \sum_k \hbar\omega_{k} \hat{a}_k^\dagger \hat{a}_k, \notag\\
	\hat{H}_{\text{vib}} &=& \sum_{k} \hbar\Omega_{k} \hat{b}_k^\dagger \hat{b}_k,\\
	\hat{H}_{\text{ph-vib}} &=& \sum_{i,j,k} \hbar g_{ijk}  \hat{q}_i \hat{q}_j \hat{Q}_k, \notag\\
	\hat{H}_{\text{vib-vib}} &=& \sum_{i,j,k} \hbar g_{ijk}^{\prime} \hat{Q}_i \hat{Q}_j \hat{Q}_k, \notag
	%+\tilde{G}_{ij}(\hat{b}_i^\dagger\hat{b}_j+\hat{b}_j^\dagger\hat{b}_i) \notag
	\label{H}
\end{eqnarray}
where $\hat{H}_{\text{ph}}$ is the energy of the phonons, which involve collective
(i.e., delocalized) hindered translation and hindered rotation (libration) of rigid molecules in a crystal.
$\hat{a}_k $ and $\hat{a}_k^{\dagger}$ are annihilation and creation operators of phonons respectively, and $\hat{q}_i = \frac{\hat{a}_i + \hat{a}^{\dagger}_i}{\sqrt{2}}$ is position operator.
Following the standard treatment, the phonon spectrum is modeled as a continuous distribution from zero frequency to the Debye cutoff frequency $\omega_{D}$, typically $100-200~\text{cm}^{-1}$
The second term $\hat{H}_{\text{vib}}$ describes the energy of vibrational modes, which involve periodic molecular deformations of individual molecules.
The doorway modes with position operator $\hat{Q}_k = \frac{\hat{b}_k + \hat{b}^{\dagger}_k}{\sqrt{2}}$, which are lower-frequency vibrations with significant bandwidths, act as a crucial bridge, where $\hat{b}_k$ and $\hat{b}^{\dagger}_k$ are annihilation and creation operators of vibrators.
They couple effectively with both lattice phonons and intramolecular vibrations, thereby channeling external energy from impact-generated phonons into the internal vibrational excitation of the molecule.
The third term $\hat{H}_{\text{ph-vib}}$ describes the conversion of continuous low-frequency phonons $\hat{q}_i$ and $\hat{q}_j$ into doorway vibrational modes $\hat{Q}_k$, and $g_{ijk}$ is the coupling coefficient.
The last term $\hat{H}_{\text{vib-vib}}$ describes the conversion between vibrational modes, with
the coupling coefficient $g_{ijk}^{\prime}$.

Under external perturbation, such as a shock, the lattice phonons (external modes) in an energetic material become excited, leading to a significant increase in their occupation numbers~\cite{kim1990theory}.
Consequently, under the mean-field approximation, i.e., $\hat{q}_i\to\langle\hat{q}_i\rangle+\delta \hat{q}_i$, the coupling from the continuum of phonon modes to the doorway vibrational modes is mapped onto an average field and a linearized interaction.
Then the phonon-vibration interaction term becomes the following (see appendix~\ref{sec:ph-vibUMA} for the detailed derivation):
\begin{eqnarray}
	\hat{H}_{\text{ph-vib}} = \sum_{k} \hbar E_{k} \hat{Q}_k+ \sum_{j,k} \hbar F_{jk}\delta \hat{q}_j \hat{Q}_k,
	\label{Hpv}
\end{eqnarray}
where the first term represents the doorway vibrational mode $\hat{Q}_k$ is subjected to the effect of an equivalent drive $E_{k}=\sum_{ij}g_{ijk}\langle\hat{q}_{i}(t)\rangle\langle\hat{q}_{j}(t)\rangle$, which can be written as
\begin{eqnarray}
    &E_{k}& = \int_{0}^{\omega_{D}}d\omega d\omega^{\prime} \frac{1}{2} G_{k}(\omega,\omega^{\prime})S(\omega)S(\omega^{\prime})  \label{Ek}\\
    && \times (\alpha(\omega)e^{-i\omega t}+\alpha^{*}(\omega)e^{i\omega t})(\alpha(\omega^{\prime})e^{-i\omega^{\prime} t}+\alpha^{*}(\omega^{\prime})e^{i\omega^{\prime} t}). \notag
\end{eqnarray}
Here, $G_{k}(\omega,\omega^{\prime})$ represents the coupling strength between two continuous phonon modes (with frequencies $\omega$ and $\omega^{\prime}$) and the $k$-th vibrational mode.
$S(\omega)$ is the state density of phonons (the continuous low-frequency phonons can be regarded as the environment of vibrational modes).
$\alpha(\omega)=\langle\hat{a}_{\omega}\rangle$ is the mean field of the phonon with frequency $\omega$.
Clearly, the collective effect of non-thermally excited phonons, generated by an initial perturbation (shock), manifests as an effective coherent drive on specific vibrational modes.
It should be emphasized that, due to interactions among phonons, phonon scattering leads to energy redistribution that can reach a quasi-equilibrium state within an extremely short time \cite{dlott2005multi,su2007correlation,michalchuk2019predicting}.
Therefore, when we refer to the non-thermal phonon population distribution induced by the initial shock, it specifically describes the distribution after phonons have attained this quasi-equilibrium state.
The second term in Eq. (\ref{Hpv}) describes the linearized interaction between phonon and vibrational mode, the coupling coefficient can be expressed as $F_{jk}=\sum_{i}2g_{ijk}\langle\hat{q}_{i}(t)\rangle$.

\begin{figure}[t]
    \includegraphics[width=7.5cm]{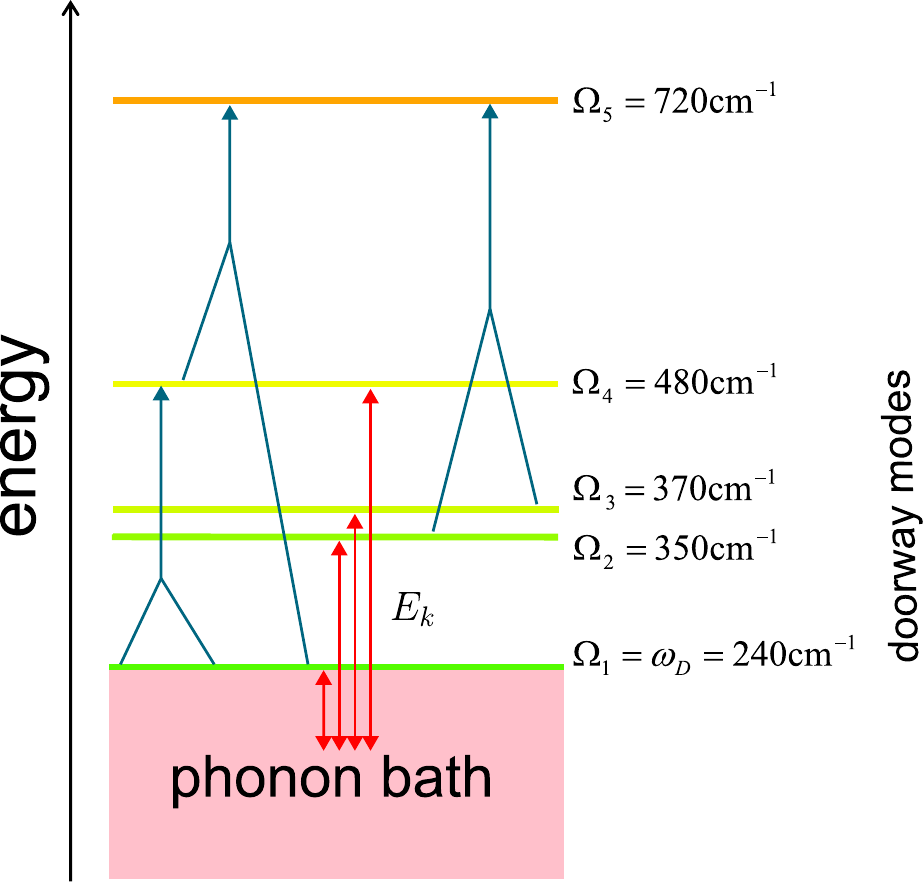}
    \caption{Schematic diagram of the simplified multiphonon up-pumping model.
	    Collective phonons extend from zero frequency to a cut off denoted $\omega_{D}$.
	    The low frequency molecular vibrations below $2\omega_{D}$ are the doorway modes ($\Omega_1, \Omega_2, \Omega_3 $ and $\Omega_4$) through which mechanical energy enters and leaves the molecule. $\Omega_5$ is  a intramolecular vibration mode, which will be excited by phonons ad doorway modes.}
    \label{Fig1}
\end{figure}

Thus, the total Hamiltonian of the system can be divided into two parts: the system composed of vibrational modes denoted by $\hat{H}_{S}$ and the environment composed of phonons denoted by $\hat{H}_{E}$,
\begin{eqnarray}
	\hat{H}_{S} &=& \sum_{k} \hbar\Omega_{k} \hat{b}_k^\dagger \hat{b}_k + \hbar E_{k} \hat{Q}_k + \sum_{i,j,k} \hbar g_{ijk}^{\prime} \hat{Q}_i \hat{Q}_j \hat{Q}_k, \notag\\
	\hat{H}_{E} &=& \sum_j \hbar\omega_{j} \hat{a}_j^\dagger \hat{a}_j, \\
	\hat{H}_{I} &=& \sum_{j,k} \hbar \frac{F_{jk}}{2} (\hat{a}_j+\hat{a}_{j}^{\dag}) (\hat{b}_k+\hat{b}_{k}^{\dag}). \notag
	\label{Hnew}
\end{eqnarray}
$\hat{H}_{I}$ is the interaction between system and environment.
In the interaction picture, the density matrix of the total system, denoted as $\hat{\rho}_{SE}$, obeys the following equation of motion:
\begin{eqnarray}
	\dot{\hat{\rho}}_{SE}(t) &=& -\frac{i}{\hbar}\left[\hat{H}_{\text{int}}(t),\hat{\rho}_{SE}(t_{i})\right]  \notag\\
	&&-\frac{1}{\hbar^{2}}\int_{t_{i}}^{t}d\tau\left[\hat{H}_{\text{int}}(t),\left[\hat{H}_{\text{int}}(\tau),\hat{\rho}_{SE}(\tau)\right]\right],
\end{eqnarray}
where $\hat{H}_{\text{int}}(t)=\sum_{j,k} \hbar \frac{F_{jk}}{2} (\hat{a}_j e^{-i\omega_{j}t}+\hat{a}_{j}^{\dag}e^{i\omega_{j}t}) (\hat{b}_k e^{-i\Omega_{k}t}+\hat{b}_{k}^{\dag}e^{i\Omega_{k}t})+\sum_{k}\hbar E_{k}\hat{Q}_k (t) + \sum_{i,j,k} \hbar g_{ijk}^{\prime} \hat{Q}_i (t) \hat{Q}_j (t) \hat{Q}_k (t)$ is the Hamiltonian in the interaction picture.
After perturbation, the initial state of the environment is not a strictly thermal state.
Under the Bonn-Markov approximation~\cite{breuer2002theory}, we can still assume that $\text{Tr}_{E}[\hat{H}_\text{int}\hat{\rho}_{S}(t_{i})\hat{\rho}_{E}(t_{i})]=0$.
Then we obtained the following Redfield equation,
% Under the Bonn-Markov approximation, the density matrix of the system satisfies the following equation:
% -\frac{i}{\hbar}\text{Tr}_{E}[\hat{H}_{I}(t),\rho_{S}(t_{i})\rho_{E}]  \notag\\
\begin{eqnarray}
	\dot{\hat{\rho}}_{S}(t) =-\frac{1}{\hbar^{2}}\text{Tr}_{E}\int_{t_{i}}^{t}d\tau\left[\hat{H}_{\text{int}}(t),\left[\hat{H}_{\text{int}}(\tau),\hat{\rho}_{S}(t)\hat{\rho}_{E}\right]\right].~~~
	\label{Redfield}
\end{eqnarray}
By tracing out the environmental degrees of freedom and applying the rotating wave approximation, then transitioning back to the Schr\"{o}dinger picture, we arrive at the following master equation (see appendix~\ref{sec:DME} for the detailed derivation),
\begin{eqnarray}
    \dot{\hat{\rho}}_S &=& -\frac{i}{\hbar} \left[ \hat{H}_{S}, \hat{\rho}_S \right] \notag\\
		       && + \sum_k \frac{\gamma_k}{2} \left( 2 \hat{b}_k \hat{\rho}_S \hat{b}_k^\dagger - \hat{b}_k^\dagger \hat{b}_k \hat{\rho}_S - \hat{\rho}_S \hat{b}_k^\dagger \hat{b}_k \right) \label{MasterE} \\
		       && + \sum_k \frac{\tilde{\gamma}_k}{2}  \left( 2 \hat{b}_k^\dagger \hat{\rho}_S \hat{b}_k - \hat{b}_k \hat{b}_k^\dagger \hat{\rho}_S - \hat{\rho}_S \hat{b}_k \hat{b}_k^\dagger \right). \notag
\end{eqnarray}
The phonon bath induced dissipation rate $\gamma_k$ and $\tilde{\gamma}_k$ can be expressed as an integral over the system-environment coupling strength, the environmental density of states, the initial environmental perturbation, and the thermal excitation number of the environment, which can be written as
\begin{eqnarray}
	\gamma_k &=& \pi\int_{0}^{\omega_{D}}d\omega d\omega^{\prime} G_{k}(\omega,\Omega_{k}-\omega) G_{k}(\omega,\omega^{\prime}) S(\omega) \notag\\
	&& \times  S(\omega^{\prime}) q(\Omega_{k}-\omega) q(\omega^{\prime}) (N_\text{th}(\omega)+1),   \label{gammak}\\
	\tilde{\gamma}_k &=& \pi\int_{0}^{\omega_{D}}d\omega d\omega^{\prime} G_{k}(\omega,\Omega_{k}-\omega) G_{k}(\omega,\omega^{\prime}) S(\omega) \notag\\
	&& \times  S(\omega^{\prime}) q(\Omega_{k}-\omega) q(\omega^{\prime}) N_\text{th}(\omega).   \label{gammawk}
\end{eqnarray}
Here, $q(\omega)=\langle\hat{q}_{\omega}\rangle$ is the mean field of the phonon with frequency $\omega$.
$N_\text{th}(\omega)=(e^{\frac{\hbar\omega}{K_{B}T}}-1)^{-1}$ is the thermal excitation number of the environment, and $K_{B}$ is Boltzmann constant, $T$ is the temperature.

Master Equation (\ref{MasterE}) describes the decoherence dynamics of doorway modes within a phonon bath after initial shock.
For high-frequency molecular vibrational modes with frequencies above $2\omega_{D}$, a similar Markovian-type decoherence channel is expected under environmental influence.
Therefore, the dynamics of such high-frequency vibrational modes can be described phenomenologically by a master equation of the same form as Eq. (\ref{MasterE}).
However, unlike the doorway modes, the high-frequency vibrational modes and the phonon modes are decoupled.
Consequently, their dissipation rate (induced by other environments) is expected to be lower than that of the doorway modes.

For comparison purposes, the phenomenological master equation of the system in the absence of external shock is presented as follows:
\begin{eqnarray}
	\dot{\hat{\rho}}_S &=& -\frac{i}{\hbar} \left[ \hat{H}_{S}^{'}, \hat{\rho}_S \right] \label{MasterE2}\\
	&& + \sum_j \frac{\kappa_j}{2}(N_\text{th}(\Omega_{j})+1) \left( 2 \hat{b}_j \hat{\rho}_S \hat{b}_j^\dagger - \hat{b}_j^\dagger \hat{b}_j \hat{\rho}_S - \hat{\rho}_S \hat{b}_j^\dagger \hat{b}_j \right)  \notag\\
	&& + \sum_j \frac{\kappa_j}{2}N_\text{th}(\Omega_{j})  \left( 2 \hat{b}_j^\dagger \hat{\rho}_S \hat{b}_j - \hat{b}_j \hat{b}_j^\dagger \hat{\rho}_S - \hat{\rho}_S \hat{b}_j \hat{b}_j^\dagger \right). \notag
\end{eqnarray}
Due to the absence of an initial shock, the driving term induced by coherent phonons in the system Hamiltonian disappears.
Thus, the system Hamiltonian $\hat{H}_{S}^{'}=\sum_{k} \hbar\Omega_{k} \hat{b}_k^\dagger \hat{b}_k + \sum_{i,j,k} \hbar g_{ijk}^{\prime} \hat{Q}_i \hat{Q}_j \hat{Q}_k$.

\begin{figure}[t]
    \includegraphics[width=0.99\linewidth]{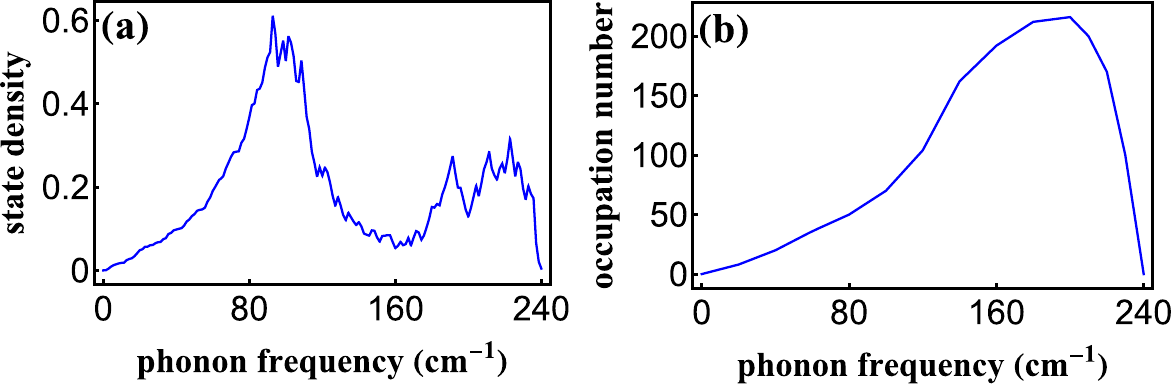}
    \caption{The distribution functions $S(\omega)$ and
	$|\alpha(\omega)|^{2}$ used in our calculations, (a) the state
	density of phonons (state/$\text{cm}^{-1}$) and (b) the
	shock-induced initial occupation number distribution of the
    phonons, where the characteristics comes from Ref.~\cite{hill1988model, kim1990theory}.}
	\label{Fig2}
\end{figure}

\section{Numerical results}
\label{sec:NR}

In this section, in order to specifically demonstrate the process of energy transfer we focus on a simplified multiphonon up-pumping model which includes four doorway modes $\Omega_{1}\sim\Omega_{4}$ and one high-frequency vibrational mode $\Omega_{5}$ (the target mode).
Schematic diagram of the energy level structure and interactions of the system is shown in Fig. \ref{Fig1}.
The Hamiltonian of the system is given by
\begin{eqnarray}
	\hat{H}_{S} &=& \sum_{k=1}^{5} \hbar\Omega_{k} \hat{b}_k^\dagger \hat{b}_k + \sum_{k=1}^{3}\hbar E_{k} \hat{Q}_k +  \hbar g_{1}^{\prime} \hat{Q}_{1}^{2} \hat{Q}_4, \notag\\
	&&+\hbar g_{2}^{\prime} \hat{Q}_1 \hat{Q}_4 \hat{Q}_5 + \hbar g_{3}^{\prime} \hat{Q}_2 \hat{Q}_3 \hat{Q}_5.
	\label{Htoy}
\end{eqnarray}
In order to determine the key parameters in the Hamiltonian or in the master equation, three key distribution functions must be known:
the coupling strength distribution function of two phonon modes that excite the $k$-th vibrational mode, denoted as $G_{k}(\omega,\omega^{\prime})$;
the state density of phonons, denoted as $S(\omega)$;
and the occupation number distribution of the phonons induced by initial shock, correspond to $|\alpha(\omega)|^{2}$.
For different molecular crystals, these distribution functions may exhibit significant differences.
In order to provide a computable example for our simplified model, we give the specific form of the distribution functions according to early study of shock-induced multiphonon up pumping in crystalline naphthalene \cite{kim1990theory,hill1988model}.

\begin{figure}[t]
	\includegraphics[width=7.5cm]{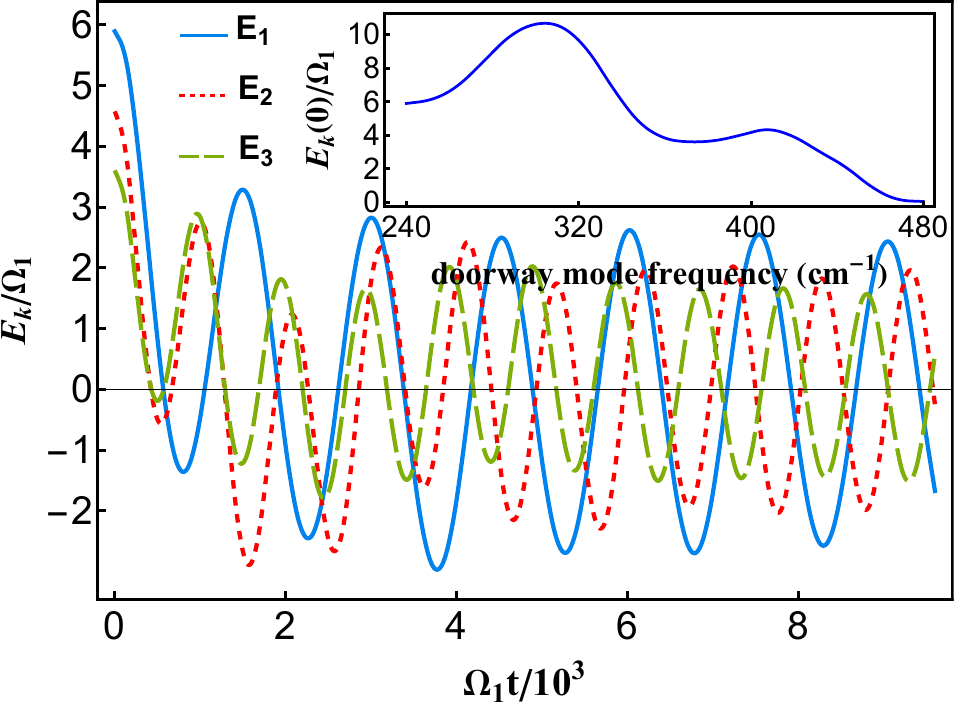}
	\caption{The long-time evolution of $E_{k}$ ($k=1,2,3$).
		Among them, the blue solid, red dotted and green dashed line represents $E_{1}$, $E_{2}$ and $E_{3}$ respectively.
		The inset figure plotting the initial value of $E_k$ for different frequency, indicate that doorway modes near $300~\text{cm}^{-1}$ may receive the strongest driving. 
		We choose $\gamma=0.36~\text{cm}^{-1}$, $d=1~\text{cm}^{-1}$. }
	\label{Fig3}
\end{figure}

We assume that the coupling strength distribution function is described by a two-dimensional Lorentzian distribution with a narrow bandwidth, i.e., $G_{k}(\omega,\omega^{\prime})=\frac{\gamma d^{2}}{(\Omega_{k}-\omega-\omega^{\prime})^{2}+d^{2}}$.
Consequently, only two-phonon processes that approximately satisfy the energy matching condition $\Omega_{k}\approx\omega+\omega^{\prime}$, can be efficiently converted into vibrational modes.
In Fig. \ref{Fig2}, we plot $S(\omega)$ and $|\alpha(\omega)|^{2}$ as a function of phonon frequency.
Based on the characteristics of the phonon density of states and the phonon occupation number provided in \cite{kim1990theory,hill1988model}, for convenience, we simulate them using random distributions.
By integrating $S(\omega)$, we can roughly estimate that the total number of phonon modes below the Debye frequency is approximately $46.2$.
In addition, compared with the low-frequency phonon mode, the shock is more likely to induce the excitation of the high-frequency mode.
By employing these distribution functions, the effective driving $E_{k}$ and the dissipation rate $\gamma_k$ and $\tilde{\gamma}_k$ can be calculated.

Integrating Eq. (\ref{Ek}), yields the time evolution of $E_{k}$ in the long-time limit, shown in Fig. \ref{Fig3}.
It is evident that the oscillation frequency of $E_{k}$ is significantly lower than that of the doorway modes.
Consequently, only the short-time dynamics of $E_{k}$ need to be considered, while its long-time anharmonic oscillations can be neglected.
The initial value of $E_{k}$ is crucial for the evolution of the system, which is illustrated in the inset of Fig. \ref{Fig3}, where $E_{k}(0)$ is plotted as a function of the doorway mode frequency.
The strongest driving occurs around a frequency of $300~\text{cm}^{-1}$, indicating that doorway modes near this frequency receive the greatest amplification.
At this point, energy can be transferred efficiently and rapidly from phonons to the doorway mode, thereby further exciting higher-frequency vibrational modes.
Achieving this requires a perfect match among three factors: the phonon spectral density of the energetic material (including coupling strength distribution and the state density of phonons), the non‑thermal distribution of phonons generated by the initial shock, and the distribution of the doorway modes.
As the frequency of the doorway mode increases further, the frequency range of the two phonons that can effectively couple with it continuously narrows.
Consequently, $E_{k}(0)$ gradually decreases and eventually approaches zero at $\Omega_{k}=480~\text{cm}^{-1}$.
This also explains why only three driving modes exist in our system, since $E_{4}\ll1$ and $E_{5}=0$.
In other words, do the candidate modes constitute genuine doorway modes that
can be effectively driven by external loading? This driving efficacy depends
critically on the phonon density of states and loading-induced redistribution
of phonon occupation numbers.
\begin{figure}[t]
    \includegraphics[width=7.5cm]{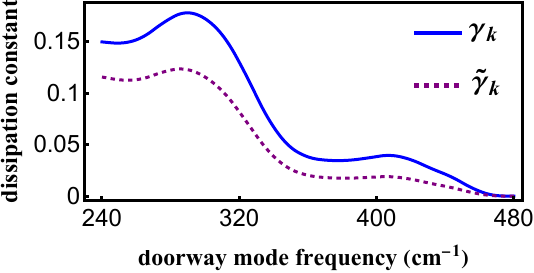}
	\caption{The dissipation rate as a function of the doorway mode frequency.
		The blue solid, purple dashed line represents $\gamma_k$ and $\tilde{\gamma}_k$ respectively.
	    The environment temperature is $400~\text{K}$. }
	\label{Fig4}
\end{figure}

On the other hand, by calculating Eqs. (\ref{gammak}) and (\ref{gammawk}), we can determine the value of the dissipation rate, i.e., $\gamma_1/\Omega_{1}\approx0.15$, $\gamma_2/\Omega_{1}\approx0.048$, $\gamma_3/\Omega_{1}\approx0.035$, $\tilde{\gamma}_1/\Omega_{1}\approx0.116$, $\tilde{\gamma}_2/\Omega_{1}\approx0.026$, $\tilde{\gamma}_3/\Omega_{1}\approx0.018$.
Similarly, in Fig.~\ref{Fig4} we plot the dissipation rate as a function of the doorway mode frequency.
The behavior closely resembles that in the inset of Fig.~\ref{Fig3}: the dissipation rate is relatively large for doorway modes around $300~\text{cm}^{-1}$, due to their strongest effective coupling with the phonon bath.
The excitation number of the corresponding doorway mode is determined by both the magnitude of the driving force and the dissipation rate.
Ideally, appropriate phonon spectral density and initial shock can lead to a high driving intensity alongside a low dissipation rate.
Beyond $300~\text{cm}^{-1}$, as the frequency of the doorway modes further increases, the dissipation rate continues to decrease and eventually approaches zero.
Here, we note that the dissipation rate induced by the phonon bath for the fourth and fifth vibrational modes are zero.
However, the system composed of these vibrational modes is not closed, in reality, each vibrational mode should be coupled to its own environment.
For simplicity, we assume that the fourth and fifth vibrational modes are coupled to another environment—distinct from the phonon bath below the Debye frequency—which gives rise to decoherence dynamics described by the master equation (\ref{MasterE2}), with corresponding dissipation rate $\gamma_4=\kappa_{4}(N_\text{th}(\Omega_{4})+1)$, $\gamma_5=\kappa_{5}(N_\text{th}(\Omega_{5})+1)$, $\tilde{\gamma}_4=\kappa_{4}N_\text{th}(\Omega_{4})$ and $\tilde{\gamma}_5=\kappa_{5}N_\text{th}(\Omega_{5})$.

\begin{figure}[t]
    \includegraphics[width=7.5cm]{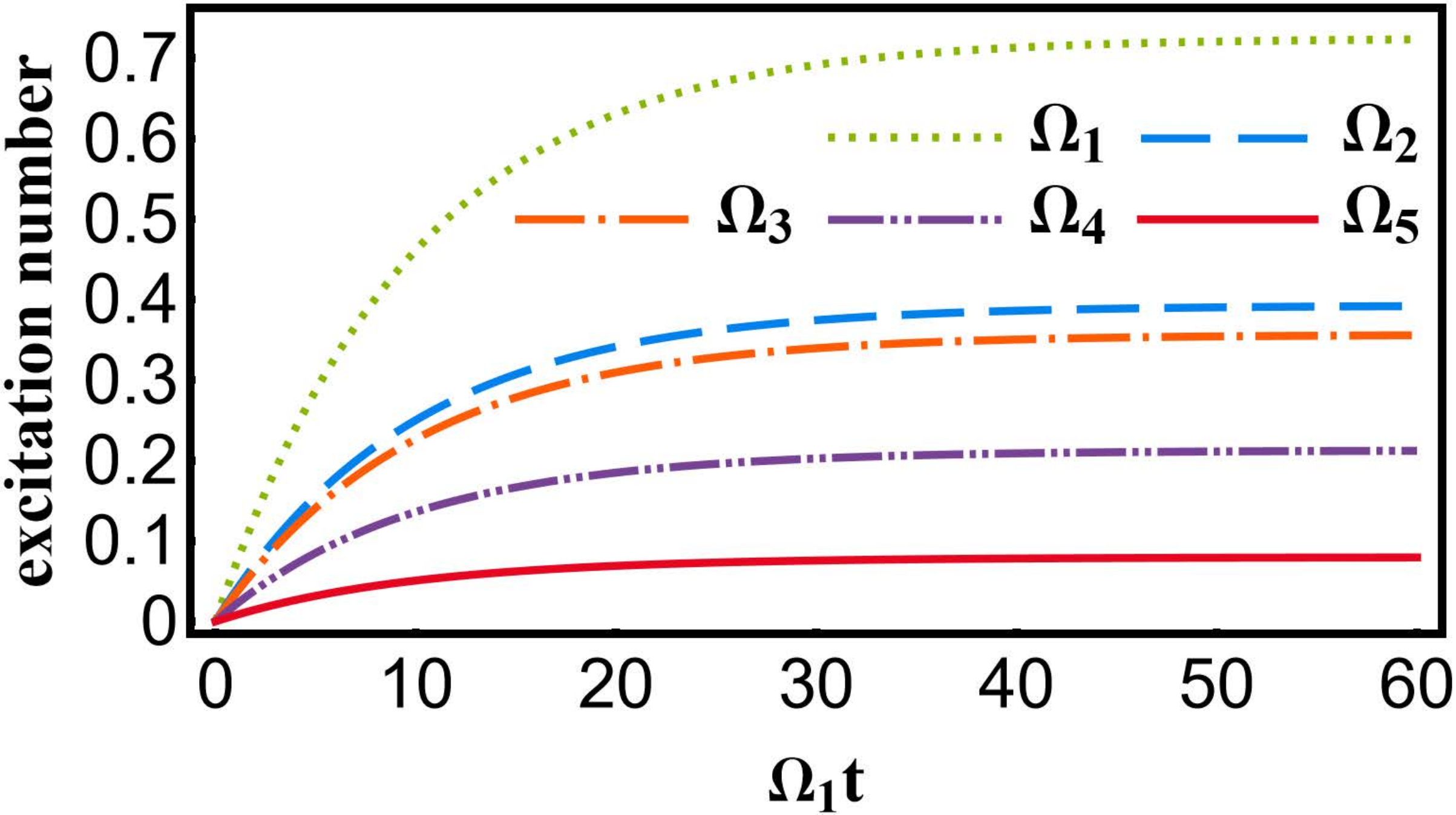}
	\caption{Thermalization dynamics of the vibrational modes.
	For convenience, we set all dissipation rates $\kappa_{j}/\Omega_{1}=0.1$, $g_{1}^{\prime}/\Omega_{1}=g_{2}^{\prime}/\Omega_{1}=0.03$, $g_{3}^{\prime}/\Omega_{1}=0.035$, and $T=400K$.}
	\label{fig:TDVM}
\end{figure}

By using the parameters calculated above, we can numerically solve the master equation (\ref{MasterE}) and (\ref{MasterE2}) to obtain the evolution of the excitation number of the vibrational mode over time.
In the absence of external shock, Fig. \ref{fig:TDVM} simulates the thermalization dynamics of the system initially at zero temperature with its environment.
The excitation numbers of the vibrational modes in the system gradually increase until thermal equilibrium with the environment is eventually established.
Due to the absence of coherent phonon pumping, the excitation of the doorway mode remains significantly suppressed, which consequently prevents the target mode $\Omega_{5}$ from being effectively populated.
If one aims to achieve excitation of the target mode through thermal equilibrium alone, the equivalent temperature of the environment would need to reach approximately $1500~\text{K}$.
Consequently, effective population of vibrational modes with even higher frequencies would require correspondingly higher environmental temperatures.

\begin{figure}[t]
    \includegraphics[width=0.85\linewidth]{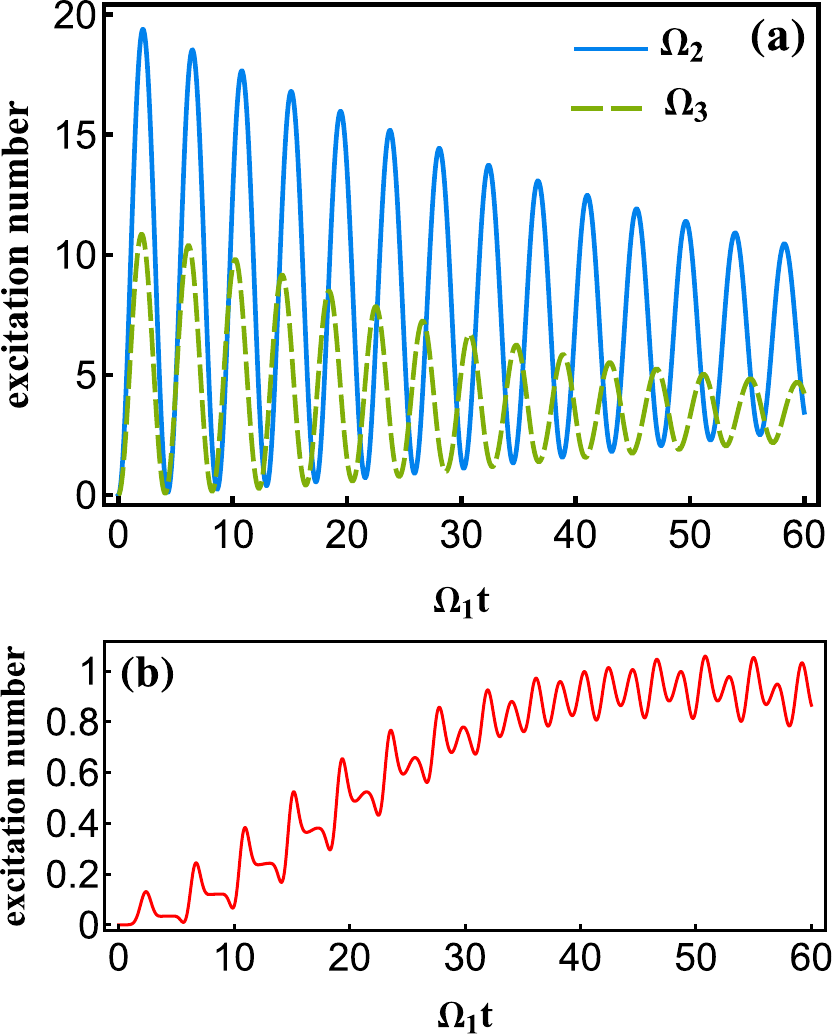}
    \caption{In the energy transfer process, two doorway modes, $\Omega_{2}$
	and $\Omega_{3}$, can effectively excite the high-frequency vibrational
        mode $\Omega_{5}$. In (a), the blue solid line and the green dashed
	line represent $\Omega_{2}$ and $\Omega_{3}$, respectively. In (b), the
	red line corresponds to $\Omega_{5}$, the targeted mode is populated significantly.  We choose
	$\kappa_{4}/\Omega_{1}=0.015$ and $\kappa_{5}/\Omega_{1}=0.005$. The
    other parameters are the same as those in Fig. \ref{fig:TDVM}.}
	\label{fig:Omega23}
\end{figure}

\begin{figure}[t]
    \includegraphics[width=0.85\linewidth]{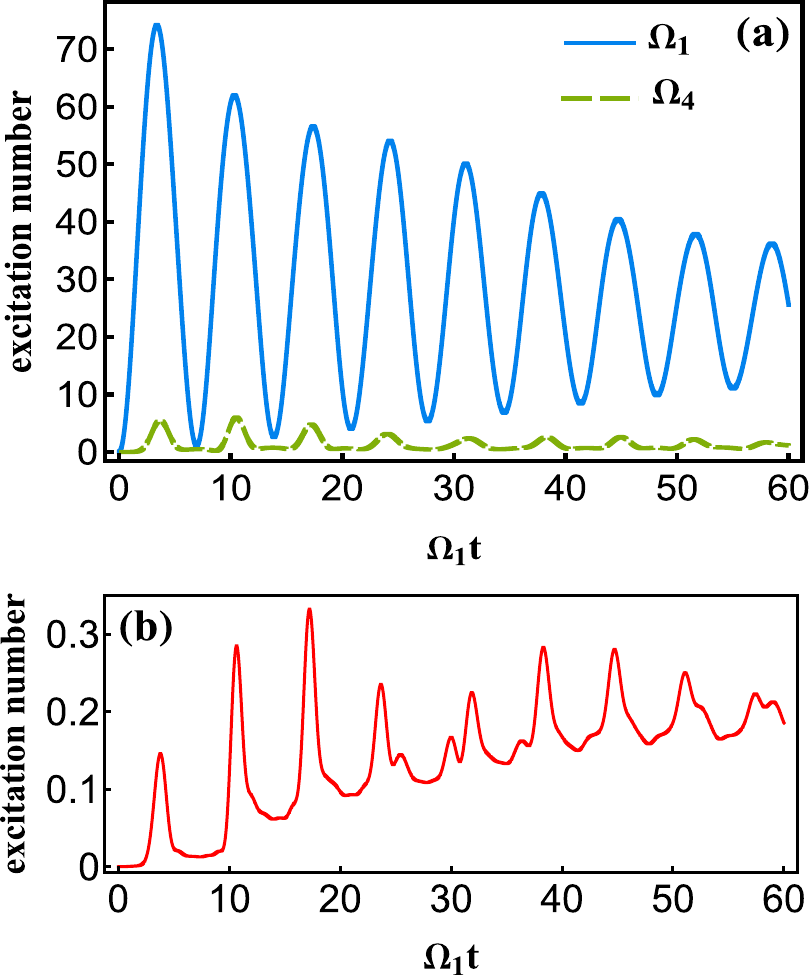}
	\caption{In the energy transfer process, two doorway modes, $\Omega_{1}$ and $\Omega_{4}$, can not effectively excite the high-frequency vibrational mode $\Omega_{5}$.
		In (a), the blue solid line and the green dashed line represent $\Omega_{1}$ and $\Omega_{4}$, respectively.
		In (b), the red line corresponds to $\Omega_{5}$.
		We choose $\kappa_{4}/\Omega_{1}=0.015$ and $\kappa_{5}/\Omega_{1}=0.005$.
		The other parameters are the same as those in Fig. \ref{fig:TDVM}.}
	\label{fig:Omega14}
\end{figure}

When a coherent phonon pumping is introduced, the situation changes markedly.
To carefully contrast the excitation of the target mode via two distinct pumping pathways, we plot the corresponding results in Figs. \ref{fig:Omega23} and \ref{fig:Omega14}, respectively.
In Fig. \ref{fig:Omega23}, only the interaction of the form $\hat{Q}_2 \hat{Q}_3 \hat{Q}_5$ is considered, while the $\hat{Q}_1 \hat{Q}_4 \hat{Q}_5$ type interaction is neglected.
In this configuration, the population of the target mode depends entirely on modes $\Omega_{2}$ and $\Omega_{3}$.
According to Fig. \ref{Fig3}, the initial drives for modes $\Omega_{2}$ and $\Omega_{3}$ are approximately $E_{2}(0)/\Omega_{1}=4.58$ and $E_{3}(0)/\Omega_{1}=3.6$, respectively.
Consequently, in panel (a), the excitation level of mode $\Omega_{2}$ is significantly higher than that of mode $\Omega_{3}$.
Under the influence of dissipation, both populations gradually stabilize.
Since both doorway modes $\Omega_{2}$ and $\Omega_{3}$ are effectively excited, the interaction of the form $\hat{Q}_2 \hat{Q}_3 \hat{Q}_5$ proceeds efficiently.
As a result, we observe in panel (b) that the target mode is successfully populated.

In Fig. \ref{fig:Omega14}, only interactions of the forms $\hat{Q}_{1}^{2} \hat{Q}_4$ and $\hat{Q}_1 \hat{Q}_4 \hat{Q}_5$ are considered, with the excitation of the target mode now governed by modes $\Omega_{1}$ and $\Omega_{4}$.
Here, the initial driving strength for mode $\Omega_{1}$ is $E_{1}(0)/\Omega_{1}=5.89$, while mode $\Omega_{4}$, being located at the boundary of the doorway mode, has no effective driving strength.
Consequently, the population of mode $\Omega_{4}$ relies primarily on the $\hat{Q}_{1}^{2} \hat{Q}_4$ interaction.
This leads to a significant disparity in the excitation levels between modes $\Omega_{1}$ and $\Omega_{4}$.
As shown in panel (a), the population of mode $\Omega_{1}$ can exceed 70, whereas that of mode $\Omega_{4}$ remains an order of magnitude lower.
Due to the weak excitation of mode $\Omega_{4}$, even the strong excitation of mode $\Omega_{1}$ is insufficient to efficiently activate the $\hat{Q}_1 \hat{Q}_4 \hat{Q}_5$ type interaction, thereby preventing effective population of the target mode.
In panel (b), we observe that during short-time simulations the population of the target mode remains around 0.3 at most, which is still far below full excitation. Comparing the Fig.~\ref{fig:Omega23} where the target mode is populated significantly, we find that multiphonon up-puming efficiency dependents whether the doorway modes coupled with target mode can be coherent driven simultaneously.

\section{Discussion and conclusion}
\label{sec:DC}

In conclusion, we have developed a fully quantum multiphonon up‑pumping model and derived the corresponding quantum master equation under the mean‑field, rotating‑wave, and Born‑Markov approximations.
Within this master equation, we have identified the origin of coherent phonon driving for doorway modes, namely the mean‑field effect of highly excited phonons induced by external shock on the vibrational modes.
Simultaneously, the phonon environment also renormalizes the dissipation experienced by the doorway modes.
Experimentally, once the post‑loading phonon excitation distribution over frequency, the phonon density of states of the material, and the two‑dimensional frequency‑dependent phonon‑vibration coupling strength are measured, the coherent driving and dissipation strengths for a vibrational mode at a given frequency can be determined quantitatively.
This allows one to identify which doorway modes are more susceptible to excitation and to further analyze the likelihood of activating targeted vibrational modes. It also provides a method to predict a EMs sensitivity to mechanical loading through calculating the effective driving $E_k$ according to the state density and initial occupation number distribution of phonons.

On the other hand, numerical simulation of the master equation itself presents a significant challenge.
In the present work, the Hilbert space of our simplified model is not excessively large, enabling us to solve the density‑matrix evolution directly in the occupation‑number representation using numerical methods such as Runge‑Kutta or linear multistep methods.
However, if the number of vibrational modes in the system is substantially increased, more advanced numerical techniques will be required, such as quantum trajectory methods~\cite{dalibard1992wave,carmichael1993open,plenio1998quantum} or stochastic Langevin equation approaches~\cite{van1992stochastic,kubo2012statistical}.
This constitutes an important direction for our future research.

\section*{ACKNOWLEDGMENTS}

This work was supported by the Research Foundation of National Key Laboratory of Shock Wave and Detonation Physics (Grant No. 2024CXPTGFJJ06407),
the National Natural Science Foundation of China (Grant No. 12304389) and the Scientific Research Foundation of NEU (Grant No. 01270021920501*115).

\appendix

\section{The phonon-vibration interaction under the meanfield approximation}
\label{sec:ph-vibUMA}

In the rotating frame with $\hat{H}_{\text{ph}}$ and under the mean-field approximation, the nonlinear interaction $\hat{H}_{\text{ph-vib}}$ in Eq. (\ref{H}) can be linearized,
\begin{eqnarray}
	&&\hat{H}_{\text{ph-vib}} \notag\\
	&&= \sum_{i,j,k} \hbar g_{ijk}  \hat{q}_i (t) \hat{q}_j (t) \hat{Q}_k   \notag\\
	&&= \sum_{i,j,k} \hbar g_{ijk} (\langle\hat{q}_i (t)\rangle+\delta \hat{q}_i (t)) (\langle\hat{q}_j (t)\rangle+\delta \hat{q}_j (t)) \hat{Q}_k  \notag\\
	&&= \sum_{i,j,k} \hbar g_{ijk} (\langle\hat{q}_i (t)\rangle \langle\hat{q}_j (t)\rangle + \langle\hat{q}_i (t)\rangle \delta \hat{q}_j (t)+ \langle\hat{q}_j (t)\rangle \delta \hat{q}_i (t)) \hat{Q}_k  \notag\\
	&&= \sum_{i,j,k} \hbar g_{ijk} \langle\hat{q}_i (t)\rangle \langle\hat{q}_j (t)\rangle \hat{Q}_k + \hbar 2 g_{ijk} \langle\hat{q}_i (t)\rangle \delta \hat{q}_j \hat{Q}_k  \notag\\
	&&= \sum_{k} \hbar E_{k} \hat{Q}_k+ \sum_{j,k} \hbar F_{jk}\delta \hat{q}_j \hat{Q}_k.
	\label{Hpvapp}
\end{eqnarray}
Here, we utilized the symmetry of the system in exchanging two phonon modes, i.e., $g_{ijk}=g_{jik}$.

\section{Derivation of the master equation}
\label{sec:DME}

The interaction Hamiltonian $\hat{H}_{\text{int}}(t)$ in the Redfield equation can be written as
\begin{eqnarray}
	\hat{H}_{\text{int}}(t) &=& \hat{H}_{e}(t) + \hat{H}_{b}(t), \label{B1} \\
	\hat{H}_{e}(t) &=& \sum_{k} \hbar \hat{\Gamma}_{k}(t)\hat{b}_{k}^{\dag}e^{i\Omega_{k}t} + \hbar \hat{\Gamma}_{k}^{\dag}(t)\hat{b}_{k}e^{-i\Omega_{k}t}, \label{B2}\\
	\hat{H}_{b}(t) &=& \sum_{k} \hbar E_{k}\hat{Q}_k (t) + \sum_{i,j,k} \hbar g_{ijk}^{\prime} \hat{Q}_i (t) \hat{Q}_j (t) \hat{Q}_k (t), ~~~~~~
\end{eqnarray}
where $\hat{\Gamma}_{k}(t)=\sum_{j} \frac{F_{jk}(t)}{2} (\hat{a}_j e^{-i\omega_{j}t}+\hat{a}_{j}^{\dag}e^{i\omega_{j}t})$ is the operators of the environment, while $\hat{H}_{b}(t)$ is the operators of the system.
We start with the following Redfield equation
\begin{equation}
	\begin{aligned}
		\dot{\hat{\rho}}_{S}(t) = & -\frac{1}{\hbar^{2}} \int_{0}^{t}d\tau \text{Tr}_{E} \Biggl(  \hat{H}_{\text{int}}(t) \hat{H}_{\text{int}}(\tau) \hat{\rho}_{S}\hat{\rho}_{E}                      \\
		                          & - \hat{H}_{\text{int}}(\tau) \hat{\rho}_{S}\hat{\rho}_{E} \hat{H}_{\text{int}}(t) -\hat{H}_{\text{int}}(t) \hat{\rho}_{S}\hat{\rho}_{E} \hat{H}_{\text{int}}(\tau) \\
		                          & + \hat{\rho}_{S}\hat{\rho}_{E} \hat{H}_{\text{int}}(\tau) \hat{H}_{\text{int}}(t) \Biggr).  \label{Redfield2}
	\end{aligned}
\end{equation}
Substituting Eq. (\ref{B1}) into Eq. (\ref{Redfield2}), and utilizing the assumption that the environment is in a thermal state (i.e., $\langle \hat{\Gamma}_{k}(t) \rangle_{th}=0$), we arrive at the following equation,
\begin{equation}
	\begin{aligned}
		\dot{\hat{\rho}}_{S}(t)  = & -\frac{1}{\hbar^{2}} \int_{0}^{t}d\tau \left[\hat{H}_{\text{b}}(t),\left[\hat{H}_{\text{b}}(\tau),\hat{\rho}_{S}\right]\right]                             \\
		                           & -\frac{1}{\hbar^{2}} \int_{0}^{t}d\tau \text{Tr}_{E} \Biggl(  \hat{H}_{\text{e}}(t) \hat{H}_{\text{e}}(\tau) \hat{\rho}_{S}\hat{\rho}_{E}                  \\
		                           & - \hat{H}_{\text{e}}(\tau) \hat{\rho}_{S}\hat{\rho}_{E} \hat{H}_{\text{e}}(t) -\hat{H}_{\text{e}}(t) \hat{\rho}_{S}\hat{\rho}_{E} \hat{H}_{\text{e}}(\tau) \\
		                           & + \hat{\rho}_{S}\hat{\rho}_{E} \hat{H}_{\text{e}}(\tau) \hat{H}_{\text{e}}(t) \Biggr).
		\label{Redfield3}
	\end{aligned}
\end{equation}
In the mean-field approximation, $\langle\hat{q}_i\rangle\gg\delta \hat{q}_i$, thus the dynamics of the system is mainly determined by $\hat{H}_{\text{b}}(t)$, which means
\begin{equation}
	\begin{aligned}
		\dot{\hat{\rho}}_{S}(t) \approx & -\frac{1}{\hbar^{2}} \int_{0}^{t}d\tau \left[\hat{H}_{\text{b}}(t),\left[\hat{H}_{\text{b}}(\tau),\hat{\rho}_{S}\right]\right]                             \\
		=                               & -\frac{i}{\hbar}\left[\hat{H}_{\text{b}}(t),\hat{\rho}_{S}\right]                                                                                          \\
		                                & -\frac{1}{\hbar^{2}} \int_{0}^{t}d\tau \text{Tr}_{E} \Biggl(  \hat{H}_{\text{e}}(t) \hat{H}_{\text{e}}(\tau) \hat{\rho}_{S}\hat{\rho}_{E}                  \\
		                                & - \hat{H}_{\text{e}}(\tau) \hat{\rho}_{S}\hat{\rho}_{E} \hat{H}_{\text{e}}(t) -\hat{H}_{\text{e}}(t) \hat{\rho}_{S}\hat{\rho}_{E} \hat{H}_{\text{e}}(\tau) \\
		                                & + \hat{\rho}_{S}\hat{\rho}_{E} \hat{H}_{\text{e}}(\tau) \hat{H}_{\text{e}}(t) \Biggr).
		\label{Redfield4}
	\end{aligned}
\end{equation}
Substituting Eq. (\ref{B2}) into Eq. (\ref{Redfield4}), we have
\begin{widetext}
	\begin{equation}
		\begin{aligned}
			\dot{\hat{\rho}}_{S}(t) = & -\frac{i}{\hbar}\left[\hat{H}_{\text{b}}(t),\hat{\rho}_{S}\right]                                                                                                                           \\
			                          & -\int_{0}^{t} d\tau \sum_{k,k'} \Bigg[
				\hat{b}_{k}^{\dagger} \hat{b}_{k'}^{\dagger} \hat{\rho}_{S}  \langle \hat{\Gamma}_{k}(t)               \hat{\Gamma}_{k'}(\tau)           \rangle  e^{i(\omega_{k} t + \omega_{k'} \tau)}
			+ \hat{b}_{k}^{\dagger} \hat{b}_{k'}           \hat{\rho}_{S}  \langle \hat{\Gamma}_{k}(t)               \hat{\Gamma}_{k'}^{\dagger}(\tau) \rangle  e^{i(\omega_{k} t - \omega_{k'} \tau)}                              \\
			                          & + \hat{b}_{k} \hat{b}_{k'}^{\dagger}           \hat{\rho}_{S}  \langle \hat{\Gamma}_{k}^{\dagger}(t)     \hat{\Gamma}_{k'}(\tau)           \rangle  e^{-i(\omega_{k} t - \omega_{k'} \tau)}
			+ \hat{b}_{k} \hat{b}_{k'}                     \hat{\rho}_{S}  \langle \hat{\Gamma}_{k}^{\dagger}(t)     \hat{\Gamma}_{k'}^{\dagger}(\tau) \rangle  e^{-i(\omega_{k} t + \omega_{k'} \tau)}                             \\
			                          & - \hat{b}_{k'}^{\dagger} \hat{\rho}_{S} \hat{b}_{k}^{\dagger}  \langle \hat{\Gamma}_{k}(t)               \hat{\Gamma}_{k'}(\tau)           \rangle  e^{i(\omega_{k} t + \omega_{k'} \tau)}
			- \hat{b}_{k'}^{\dagger} \hat{\rho}_{S} \hat{b}_{k}            \langle \hat{\Gamma}_{k}^{\dagger}(t)     \hat{\Gamma}_{k'}(\tau)           \rangle  e^{-i(\omega_{k} t - \omega_{k'} \tau)}                             \\
			                          & - \hat{b}_{k'}           \hat{\rho}_{S} \hat{b}_{k}^{\dagger}  \langle \hat{\Gamma}_{k}(t)               \hat{\Gamma}_{k'}^{\dagger}(\tau) \rangle  e^{i(\omega_{k} t - \omega_{k'} \tau)}
			- \hat{b}_{k'}           \hat{\rho}_{S} \hat{b}_{k}            \langle \hat{\Gamma}_{k}^{\dagger}(t)     \hat{\Gamma}_{k'}^{\dagger}(\tau) \rangle  e^{-i(\omega_{k} t + \omega_{k'} \tau)}                             \\
			                          & - \hat{b}_{k}^{\dagger} \hat{\rho}_{S} \hat{b}_{k'}^{\dagger}  \langle \hat{\Gamma}_{k'}(\tau)           \hat{\Gamma}_{k}(t)               \rangle  e^{i(\omega_{k} t + \omega_{k'} \tau)}
			- \hat{b}_{k}^{\dagger} \hat{\rho}_{S} \hat{b}_{k'}            \langle \hat{\Gamma}_{k'}^{\dagger}(\tau) \hat{\Gamma}_{k}(t)               \rangle  e^{i(\omega_{k} t - \omega_{k'} \tau)}                              \\
			                          & - \hat{b}_{k}           \hat{\rho}_{S} \hat{b}_{k'}^{\dagger}  \langle \hat{\Gamma}_{k'}(\tau)           \hat{\Gamma}_{k}^{\dagger}(t)     \rangle  e^{-i(\omega_{k} t - \omega_{k'} \tau)}
			- \hat{b}_{k}           \hat{\rho}_{S} \hat{b}_{k'}            \langle \hat{\Gamma}_{k'}^{\dagger}(\tau) \hat{\Gamma}_{k}^{\dagger}(t)     \rangle  e^{-i(\omega_{k} t + \omega_{k'} \tau)}                             \\
			                          & + \hat{\rho}_{S} \hat{b}_{k'}^{\dagger}  \hat{b}_{k}^{\dagger} \langle \hat{\Gamma}_{k'}(\tau)           \hat{\Gamma}_{k}(t)               \rangle  e^{i(\omega_{k} t + \omega_{k'} \tau)}
			+ \hat{\rho}_{S} \hat{b}_{k'}^{\dagger} \hat{b}_{k}            \langle \hat{\Gamma}_{k'}(\tau)           \hat{\Gamma}_{k}^{\dagger}(t)     \rangle  e^{-i(\omega_{k} t - \omega_{k'} \tau)}                             \\
			                          & + \hat{\rho}_{S} \hat{b}_{k'}  \hat{b}_{k}^{\dagger}           \langle \hat{\Gamma}_{k'}^{\dagger}(\tau) \hat{\Gamma}_{k}(t)               \rangle  e^{i(\omega_{k} t - \omega_{k'} \tau)}
				+ \hat{\rho}_{S} \hat{b}_{k'}  \hat{b}_{k}                     \langle \hat{\Gamma}_{k'}^{\dagger}(\tau) \hat{\Gamma}_{k}^{\dagger}(t)     \rangle  e^{-i(\omega_{k} t + \omega_{k'} \tau)} \Bigg].
			\label{Redfield5}
		\end{aligned}
	\end{equation}
\end{widetext}
By employing the rotating-wave approximation, in which rapidly oscillating terms are neglected, and further assuming that the correlation time of the bath is sufficiently short so that the upper limit of the integration can be extended to infinity, Eq. (\ref{Redfield5}) can be reduced to the following form
\begin{equation}
	\begin{aligned}
		\dot{\hat{\rho}}_{S}(t) = & -\frac{i}{\hbar}\left[\hat{H}_{\text{b}}(t),\hat{\rho}_{S}\right]                                                                                     \\
		                          & - \int_{0}^{\infty} d\tau \sum_{k} \Bigg[
		\hat{b}_{k}^{\dagger} \hat{b}_{k} \hat{\rho}_{S} \langle \hat{\Gamma}_{k}(\tau) \hat{\Gamma}_{k}^{\dagger}(0) \rangle  e^{i\omega_{k} \tau}                                       \\
		                          & + \hat{b}_{k} \hat{b}_{k}^{\dagger} \hat{\rho}_{S} \langle \hat{\Gamma}_{k}^{\dagger}(\tau) \hat{\Gamma}_{k}(0) \rangle  e^{-i\omega_{k} \tau}        \\
		                          & - \hat{b}_{k}^{\dagger} \hat{\rho}_{S} \hat{b}_{k} \langle \hat{\Gamma}_{k}^{\dagger}(\tau) \hat{\Gamma}_{k}(0) \rangle  e^{-i\omega_{k} \tau}        \\
		                          & - \hat{b}_{k}^{\dagger} \hat{\rho}_{S} \hat{b}_{k} \langle \hat{\Gamma}_{k}^{\dagger}(0) \hat{\Gamma}_{k}(\tau) \rangle  e^{i\omega_{k} \tau}         \\
		                          & - \hat{b}_{k} \hat{\rho}_{S} \hat{b}_{k}^{\dagger} \langle \hat{\Gamma}_{k}(\tau) \hat{\Gamma}_{k}^{\dagger}(0) \rangle  e^{i\omega_{k} \tau}         \\
		                          & - \hat{b}_{k} \hat{\rho}_{S} \hat{b}_{k}^{\dagger} \langle \hat{\Gamma}_{k}(0) \hat{\Gamma}_{k}^{\dagger}(\tau) \rangle  e^{-i\omega_{k} \tau}        \\
		                          & + \hat{\rho}_{S} \hat{b}_{k}^{\dagger} \hat{b}_{k} \langle \hat{\Gamma}_{k}(0) \hat{\Gamma}_{k}^{\dagger}(\tau) \rangle  e^{-i\omega_{k} \tau}        \\
		                          & + \hat{\rho}_{S} \hat{b}_{k} \hat{b}_{k}^{\dagger} \langle \hat{\Gamma}_{k}^{\dagger}(0) \hat{\Gamma}_{k}(\tau) \rangle  e^{i\omega_{k} \tau} \Bigg].
		\label{Redfield6}
	\end{aligned}
\end{equation}
By introducing the dissipation rate $\gamma_k$ and $\tilde{\gamma}_k$ as well as the Lamb and Stark shift, which satisfy
\begin{eqnarray}
	\int_{0}^{\infty} d\tau \langle \hat{\Gamma}_{k}(\tau) \hat{\Gamma}_{k}^{\dagger}(0) \rangle  e^{i\omega_{k} \tau}  &=& \frac{\gamma_k}{2} + i \delta_{k},  \\
	\int_{0}^{\infty} d\tau \langle \hat{\Gamma}_{k}(0) \hat{\Gamma}_{k}^{\dagger}(\tau) \rangle  e^{-i\omega_{k} \tau} &=& \frac{\gamma_k}{2} - i \delta_{k},  \\
	\int_{0}^{\infty} d\tau \langle \hat{\Gamma}_{k}^{\dagger}(0) \hat{\Gamma}_{k}(\tau) \rangle  e^{i\omega_{k} \tau}  &=& \frac{\tilde{\gamma}_k}{2} + i \varepsilon_{k},  \\
	\int_{0}^{\infty} d\tau \langle \hat{\Gamma}_{k}^{\dagger}(\tau) \hat{\Gamma}_{k}(0) \rangle  e^{-i\omega_{k} \tau} &=& \frac{\tilde{\gamma}_k}{2} - i \varepsilon_{k}.
\end{eqnarray}
Back to Schr\"{o}dinger picture, as the Lamb and Stark shift is extremely small compared to the system's Hamiltonian, so it can be ignored, then we obtain the master equation of the vibrational modes for the multiphonon up-pumping model:
\begin{eqnarray}
	\dot{\hat{\rho}}_S &=& -\frac{i}{\hbar} \left[ \hat{H}_{S}, \hat{\rho}_S \right] \notag\\
	&& + \sum_k \frac{\gamma_k}{2} \left( 2 \hat{b}_k \hat{\rho}_S \hat{b}_k^\dagger - \hat{b}_k^\dagger \hat{b}_k \hat{\rho}_S - \hat{\rho}_S \hat{b}_k^\dagger \hat{b}_k \right) \\
	&& + \sum_k \frac{\tilde{\gamma}_k}{2}  \left( 2 \hat{b}_k^\dagger \hat{\rho}_S \hat{b}_k - \hat{b}_k \hat{b}_k^\dagger \hat{\rho}_S - \hat{\rho}_S \hat{b}_k \hat{b}_k^\dagger \right). \notag
\end{eqnarray}
The detailed derivation of the dissipation rate $\gamma_k$ can be presented as follows,
\begin{eqnarray}
	\gamma_k &=& 2 \mathrm{Re} \left( \int_{0}^{\infty} d\tau \langle \hat{\Gamma}_{k}(\tau) \hat{\Gamma}_{k}^{\dagger}(0) \rangle  e^{i\omega_{k} \tau} \right)  \notag\\
	&=& \mathrm{Re} \Biggl( \int_{0}^{\infty} d\tau \sum_{ij} \frac{F_{ik}(\tau)F_{jk}^{*}(0)}{2} \langle (\hat{a}_{i} e^{-i\omega_{i}\tau} + \hat{a}_{i}^{\dagger} e^{i\omega_{i}\tau} ) \notag\\
	&& \times ( \hat{a}_{j} + \hat{a}_{j}^{\dagger} ) \rangle e^{i\omega_{k}\tau} \Biggr).
\end{eqnarray}
As the environment is in a thermal state, then
\begin{eqnarray}
	\gamma_k &=& \mathrm{Re} \Biggl( \int_{0}^{\infty} d\tau \sum_{j} \frac{F_{jk}(\tau)F_{jk}^{*}(0)}{2} (\langle \hat{a}_{j} \hat{a}_{j}^{\dagger} \rangle e^{i(\omega_{k}-\omega_{j})\tau} \notag\\
		&& + \langle \hat{a}_{j}^{\dagger} \hat{a}_{j} \rangle e^{i(\omega_{k}+\omega_{j})\tau}) \Biggr) \notag\\
	&=& \mathrm{Re} \Biggl( \int_{0}^{\infty} d\tau \sum_{jlm} g_{ljk} g_{mjk}^{*} (\alpha_{m}+\alpha_{m}^{*}) \\
	&& \times (\alpha_{l} \langle \hat{a}_{j} \hat{a}_{j}^{\dagger} \rangle e^{i(\omega_{k}-\omega_{j}-\omega_{l})\tau} + \alpha_{l} \langle \hat{a}_{j}^{\dagger} \hat{a}_{j} \rangle e^{i(\omega_{k}+\omega_{j}-\omega_{l})\tau}  \notag\\
	&& + \alpha_{l}^{*} \langle \hat{a}_{j} \hat{a}_{j}^{\dagger} \rangle e^{i(\omega_{k}-\omega_{j}+\omega_{l})\tau} + \alpha_{l}^{*} \langle \hat{a}_{j}^{\dagger} \hat{a}_{j} \rangle e^{i(\omega_{k}+\omega_{j}+\omega_{l})\tau}) \Biggr). \notag
\end{eqnarray}
Here $\alpha_{l}=\langle \hat{a}_{l} \rangle$ is the mean field of the phonons.
Applying the rotating wave approximation once more and accounting for the energy-matching condition in the multi‑phonon up-pumping process, namely, that the frequency of the $k$‑th vibrational mode equals the sum of the frequencies of the $j$‑th and $l$‑th phonons,
then we find
\begin{eqnarray}
	\gamma_k &=& \mathrm{Re} \Biggl( \int_{0}^{\infty} d\tau \sum_{jlm} g_{ljk} g_{mjk}^{*} (\alpha_{m}+\alpha_{m}^{*}) \notag\\
	&& \times \alpha_{l} (\langle \hat{N}_{j} \rangle_{th} +1) e^{i(\omega_{k}-\omega_{j}-\omega_{l})\tau} \Biggr) \notag\\
	&=&  \mathrm{Re} \Biggl( \pi \sum_{jlm} g_{ljk} g_{mjk}^{*} (\alpha_{m}+\alpha_{m}^{*})  \notag\\
	&& \times \alpha_{l} (\langle \hat{N}_{j} \rangle_{th} +1) \delta(\omega_{k}-\omega_{j}-\omega_{l}).
\end{eqnarray}
Transform the summation into an integral, and eliminate an integral variable by using the $\delta$ function, we finally obtained the following equation
\begin{eqnarray}
	\gamma_k &=& \pi\int_{0}^{\omega_{D}}d\omega d\omega^{\prime} G_{k}(\omega,\omega_{k}-\omega) G_{k}(\omega,\omega^{\prime}) S(\omega) \notag\\
	&& \times  S(\omega^{\prime}) q(\omega_{k}-\omega) q(\omega^{\prime}) (\langle\hat{N}(\omega)\rangle_{th}+1).
\end{eqnarray}
Here, $\omega_{k}$ represents the doorway mode $\Omega_{k}$, and similarly, the expression of $\tilde{\gamma}_k$ can also be derived using the same method.

%%%%%%%%%%%%%%%%%%%%%%%%%%%%%%%%%%%%%%%%%%%%%%%%%%%%%%%%%%%%%%%%%%%%%%%%%%%%%%%%%%%%%%%%%%%%%%%%%%%%%%%%%%%%%%%%%%%%%%%%%

\bibliography{ref}
\end{document}